\documentstyle{article}
\title{\bf Closed timelike curves in superfluid $^{3}$He}
\author{Pedro F. Gonz\'alez-D\'{\i}az and Carmen L. Sigenza\\
Instituto de Matem\'aticas y F\'{\i}sica Fundamental\\
Consejo Superior de Investigaciones Cient\'{\i}ficas\\
Serrano 121, 28006 Madrid (SPAIN)\\
}
\date{June 15, 1998}
\begin{document}
\maketitle
\large
\setlength{\baselineskip}{0.9cm}
%\vspace{3cm}

%\renewcommand{\theequation}{\thesection . \arabic{equation} }

\begin{center}
{\bf Abtract}
\end{center}

\noindent It is shown that the curved spacetime induced in
a thin film of superfluid
$^{3}$He-A by the presence of symmetric vortices with
the unbroken symmetry
phase, admits the existence of closed timelike curves through
which only superfluid clusters formed by anti-$^{3}$He atoms
can travel and violate causality.

\vspace{.5cm}


\vspace{.3cm}


\vspace{1.5cm}

\noindent Corresponding Author: P.F. Gonz lez-D¡az,
Instituto de Matem ticas y F¡sica Fundamental,
C.S.I.C., Serrano 121, 28006-Madrid, Spain

\noindent Telephon: (34-1)5616800 \hspace*{.3cm} Fax: (34-1)5645557

\noindent E-Mail: iodpf21@cc.csic.es

\pagebreak

One of the most appealing recent developments in physics is
the discovery of a common boundary between condensed-matter
systems at low temperatures and some cosmological and gravitational
systems [1,2]. It has been theoretically established [3] and
experimentally checked [4] that the vortices which form
when $^{3}$He-A undergoes
the phase transition to its superfluid phase [5] can be
regarded as the condensed matter analogs
of the so called cosmic strings (i.e. the gauge-theory
topological defects which are thought to have been created
through phase transitions in the early stages of the
evolution of the universe and that could well have been
the seeds for presently observed galaxies [6]). On the
other hand, it has also been pointed out [7] that event
horizons and ergoregions similar to those occurring in
black and white holes with angular momentum can be
found as well in the curved spacetime of
planar solitons moving in superfluid
$^{3}$He-A.

Most symmetric vortices in $^{3}$He-A can be created in
thin films where the unit vector $\hat{\ell}$ defining
the direction of the gap nodes in momentum space is
fixed along the normal to the film, and the superfluid
circulates around the vortex axis with a velocity
given by [8]
\begin{equation}
v_s=\frac{\hbar}{2mr} ,
\end{equation}
with $m$ the mass of the $^{3}$He atoms and $r$ the
distance to the vortex axis. In this case, close to
their two zeros, the energy spectrum of the
Bogoliubov-Nambu fermion quasiparticles becomes that
of a charged, massless relativistic particle
propagating in a curved spacetime whose cylindrically
symmetric metric can be written as [9]:
\begin{equation}
ds^2=\left(c_{\perp}^2 -v_s^2\right)dt^2
+\left(\frac{\hbar^2}{4m^2 c_{\perp}^2}-r^2\right)
\left(1-\frac{v_s^2}{c_{\perp}^2}\right)^{-1}d\phi^2
-\frac{c_{\perp}^2 dz^2}{c_{\parallel}^2}-dr^2
+\frac{\hbar}{m}dtd\phi ,
\end{equation}
where $c_{\perp}=\frac{\bigtriangleup}{p_F}$ and
$c_{\parallel}=v_F=\frac{k_F}{m}$ are the speed of
light transverse to and along $\hat{\ell}$, respectively,
with ${\bf k}_F=P_F \hat{\ell}$ the Fermi momentum and
$\bigtriangleup << k_F v_F$ the gap amplitude. In metric
(2) the true singularity at the horizon $v_s=c_{\perp}$
marks the onset of the ergoregion with $v_s>c_{\perp}$
and occurs at exactly the radius of the vortex core:
\begin{equation}
r_c=\frac{\hbar}{2mc_{\perp}} .
\end{equation}

In this letter we shall consider the possible formation
of closed timelike curves (CTCs) in the spacetime
decribed by metric (2). It will be seen that ordinary
$^{3}$He atoms in the superfluid phase cannot enter
any of such CTCs and that it is only for $^{3}$He
anti-atoms, which are made of the antiparticles to the
particles that form up an $^{3}$He atom, that the
spacetime described by metric (2) can display CTCs.
We also discuss the possibility to make observable the
effects that such CTCs might have in the neighbourhood
of the transition temperature to the superfluid phase.

We first note that the line element (2) becomes the
spacetime metric of a spinning cosmic string [10] at
points sufficiently far from the vortex axis,
$r>>r_c$, only if we interpret the quantity $\hbar/8mG$ as
the internal angular momentum $J$ per unit
length in the resulting spinning string and adscribe
to this a zero mass per unit length, $\mu=0$.
Far from the vortex axis, metric (2) reduces to
\begin{equation}
ds^2=c_{\perp}^2 dt^2
+\left(\frac{\hbar^2}{4m^2 c_{\perp}^2}-r^2\right)d\phi^2
-\frac{c_{\perp}^2 dz^2}{c_{\parallel}^2}-dr^2
+\frac{\hbar}{m}dtd\phi .
\end{equation}
Much as for the
cylindrically symmetric
metric that describes the spacetime of
a homogeneously rotating G"del
universe [11], at first glance, one could expect
that CTCs would be formed in some regions of the spacetime
described by metric (4). However, in order for the Killing
vector $\partial_{\phi}$ with closed orbits
to be timelike everywhere and so
allow for CTCs, the coordinate $r$ must be constrained to
the range $r<r_c$, a condition that clearly contradicts the
approximation where metric (4) is valid.

For the more general metric (2) one would in principle not
expect CTCs to appear, too, neither outside nor inside the
horizon at $r_c$, the reason for the latter case being that
the condition $r<r_c$ necessarily implies $v_s >c_{\perp}$,
with $v_s=c_{\perp}$ at $r=r_c$, and
hence the Killing vector $\partial_{\phi}$ keeps being
spacelike even on the region inside the vortex. Let us
consider however the line $L_0$ defined by
$L_0 : t=+\beta\phi, r=R, z=0$. For that line the
$g_{\phi\phi}$-component of the metric tensor becomes:
\begin{equation}
c_{\perp}^{-2}g_{\phi\phi}=
\frac{\hbar^2}{4m^2 c_{\perp}^2}-r^2
+\frac{\hbar\beta}{mc_{\perp}^2}\left(c_{\perp}^2-v_s^2\right).
\end{equation}
Thus, $\partial_{\phi}$ can still be timelike, provided that
\begin{equation}
r_c < r=R < r_0=\sqrt{\frac{\hbar\beta}{m}},
\end{equation}
with
\begin{equation}
\beta > \frac{\hbar}{4mc_{\perp}^2} > 0 .
\end{equation}

We note now that any $^{3}$He atoms propagating along any line
satisfying condition (6), (7) forward in time can also be described
as the corresponding anti-$^{3}$He atoms (i.e. atoms that are
made of the antiparticles to the particles which an $^{3}$He
atom is made of) moving backward in time on that line.
Thus, one can
always consider another line in the spacetime described by
metric (2), defined as
$L_1 : t=-\alpha\phi, r=R, z=0$, along which anti-$^{3}$He
atoms (with negative mass -$m$) propagate backward in time.
If we set
$\alpha$ and $\beta$ to be both positive and satisfying
$\alpha >\beta$, then line $L_1$ will also be timelike
everywhere and, since an initial point $Q(\phi=0)$ and a
final point $P(\phi=2\pi)$ on $L_1$ will also be on $L_0$,
where $P$ precedes $Q$, one can readily deduce that,
relative to a given observer
that evolves forward in time,
CTCs are actually allowed to occur in the exterior superfluid
region $r>r_c$ of the
spacetime described by metric (2), provided that
superfluid $^{3}$He can travel through
such CTCs
only in the form of anti-atoms.

Nevertheless, the probability for these CTCs to exist and carry with
superfluid clusters of atoms through them would not depend
on the presence of some nonzero proportion of anti-$^{3}$He
or any pair creation process originally in the sample.
Relative to a given observer evolving forward in time,
even when no initial trace of
anti-$^{3}$He existed, the above CTCs had to exist and be
operative, since ordinary $^{3}$He atoms traveling through
them would behave like their corresponding anti-atoms to
the given observer.

Finally, we briefly comment on the possibility of an
experimental verification of these CTCs. This would naturally
reduce to the question, how could superfluid anti-$^{3}$He
be detected in an experiment with normal superfluid $^{3}$He?.
In order to try to answer this question, let us suppose that
anti-$^{3}$He is traveling through a CTC into the past.
Of course, it can do so by time amounts which depend on the
value of $\beta$ and, therefore, such a traveling can be
so large as for making an observer able to discriminate
the presence of superfluid anti-$^{3}$He in supercooling
experiments on thin films containing ultrapure (without
any trace of $^{4}$He) $^{3}$He, before this has reached
the superfluid transition temperature $T_c$. If superfluid
anti-$^{3}$He can travel into the past the way we have
discussed in this letter, one could expect that before
reaching $T_c$ clusters of some tens of anti-$^{3}$He
atoms able to show superfluid behaviour would momentary
appear still in the viscous $^{3}$He phase, even though
the sample did not contain any trace of $^{4}$He.

A way to try to detect these superfluid anti-$^{3}$He
clusters would require dissolving a given proportion of
a suitable small molecule $A$ in the liquid helium,
checking whether the added molecules are able to freely
rotate in short time intervals by the use of ultrafast
laser spectroscopy, in an experiment similar to that
recently performed by Grebenev, Toennies and Vilesov to
determine the minimum number of helium atoms needed for
superfluidity [12]. However, expectation to detect CTCs in
this way would be very small because viscous $^{3}$He
should attract the molecules much more strongly than
anti-$^{3}$He would do. This expectation very much
increased if, instead of molecule A, we were able to
use the corresponding anti-molecule around which
anti-$^{3}$He atoms could much easierly cluster.

Although superfluid liquid helium acts as a true vacuum
with respect to its viscous phase,
experiments like the one discussed above would also
be useful to check
violation of causality because of the presence of an observer
who could always decide to stop and invert the supercooling
process before it reached the transition temperature $T_c$, while
still detecting causally-disconnected clusters of superfluid
anti-$^{3}$He in the viscous liquid.

\vspace*{.5cm}

\noindent {\bf Acknowledgements}.
The authors acknowledge DGICYT by support under Research Project
No. PB94-0107, and A. Ferrera, L.J. Garay
and G.A. Mena Marug n for useful comments.

\pagebreak

\noindent\section*{References}
\begin{description}
\item [1] W.H. Zurek, Phys. Rep. 276 (1997) 177.
\item [2] G.E. Volovik and T. Vachaspati, Int. J. Mod. Phys. B10,
(1996) 471.
\item W.H. Zurek, Nature 317 (1985) 505.
\item [4] P.C. Hendry, N.S. Lawson, R.A.M. Lee, P.V.E. McClintock
and C.H.D. Williams, Nature 386 (1994) 315; C. B"uerle, Yu.M.
Bunkov, S.N. Fisher, H. Godfrin and G.R. Pickett, Nature 382
(1996) 332; V.M.H. Ruutu, V.B. Eltsov, A.J. Gill, T.W.B. Kibble,
M. Krusius, Yu.G. Makhlin, B. Placais, G.E. Volovik and W. Su,
Nature 382 (1996) 334.
\item [5] A.J. Leggett and S. Yip, in: {\it Helium Three}, eds.
W.P. Halperin and L.P. Pitaevskii (Elsevier Science Publishers,
B.V., 1990).
\item [6] N. Turok and R.H. Brandenberger, Phys. Rev. D33 (1986) 2175.
\item [7] T.A. Jacobson and G.E. Volovik, {\it Event horizons
and ergoregions in $^{3}$He}, cond-mat/9801308; W.G. Unruh,
Phys. Rev. Lett. 46 (1981) 1351; D51 (1995) 2827.
\item [8] G.E. Volovik, {\it Superfluid $^{3}$He, Particle
Physics and Cosmology}, cond-mat/9711031.
\item [9] G.E. Volovik, Pis'ma ZhETF 62 (1995) 58 [Engl.
transl.: JETP Lett. 62 (1995) 65].
\item [10] D. Harari and A.P. Polychronakos, Phys. Rev. D38
(1988) 3320.
\item [11] K. G"del, Rev. Mod. Phys. 21 (1949) 447.
\item [12] S. Grebenev, J.P. Toennies and A. Vilesov,
Science 279 (1998) 2083.

\end{description}

\end{document}